\documentstyle[pra,aps]{revtex}
\begin{document}
\draft
\title{Understanding Quantum Entanglement of Bipartite
System Based on Bell Inequality }
\author{Li-Bin Fu$^1$
\footnote{Email address: fu$_-$libin@mail.iapcm.ac.cn},
Jing-Ling Chen$^{1,2}$
\footnote{Email address: jinglingchen@eyou.com},
Xian-Geng Zhao$^1$, and Shi-Gang Chen$^1$}
\address{$^1$Institute of Applied Physics and Computational Mathematics,\\
P.O. Box 8009 (26), 100088 Beijing, China\\
$^2$Department of Physics, Faculty of Science, National University
of Singapore, \\ Lower Kent Ridge, Singapore 119260, Republic of
Singapore.}
\maketitle

\begin{abstract}
\begin{center}
{\bf Abstract}
\end{center}

We present an alternative definition of quantum entanglement for
bipartite system based on Bell inequality and operators'
noncommutativity. A state is said to be entangled, if the maximum
of CHSH expectation value $F_{\max }$ is obtain by noncommutative
measures on each particle of the bipartite system; otherwise, the
state is a disentangled state. A uniform measure quantifying the
degree of entanglement for any state of the bipartite system is also
proposed.

\end{abstract}


The importance of quantum entanglement and Bell inequality is not
extravagant. Actually, it has been stressed in voluminous articles
that the essential feature of quantum mechanics that is taken
advantage of by quantum computation is {\it quantum entanglement},
the phenomenon referred to by Schr\"odinger as ``the
characteristic trait of quantum mechanics'' \cite{sch}. In recent
year, quantum entanglement has become not only a tool for exposing
the weirdness of quantum mechanics \cite{bell,epr,chsh}, but also
a more powerful resource in a number of applications, such as
quantum cryptography \cite{myd2,gmdr}, teleportation \cite{myd4},
dense coding \cite{myd3}, communication protocols and
computation \cite{myd6,myd7}. The superior potentiality of
entangled states has raised a nature question: ``How much are two
particles entangled?'' As a foundation property of quantum system,
quantum entanglement has been extensively studied in connection
with Bell inequality \cite{bell}. For a pure state of a two-qubit
system, because there exists an equivalence between ``entangled
states'' and ``states that violate the Bell
inequality''\cite{gqn2,gqn3}, it is not difficult to quantify the
degree of entanglement for two qubits in a pure state. Much has
been done in this field \cite {scd,myd9,depure} and all of them
are consistent with the same physical interpretation: how much a
pure state violates the Bell inequality. However, theory of
entanglement for mixed quantum states is much more complicated and
still less well understood. Even qualitative distinctions between
local and nonlocal states are unclear, for example, many entangled
mixed states never violate the Bell inequality. It seems that people
still lack
witness to determine whether a mixed states is entangled. Although
there are many works devoted to quantify the entanglement of mixed
states \cite{myd9,h3,jens}, the physical interpretations of them
are somewhat different from one another.

The difficulty we encounter for mixed states is that the
equivalence between ``entangled states'' and ``states that violate
the Bell inequality'' is no longer held. Nevertheless, as a
foundation for testing nonlocal quantum correlation of
states, we believe that Bell inequality is a powerful tool to
determine the nonlocal correlations for not only pure states, but
also mixed states. An important and essential difference between
quantum and classical theory is the commutativity of operators,
namely, the quantum measurement can be noncommutative, while the
classical measurement must be commutative. Indeed, the
relationship between the noncommutativity of operators and the
violation of the Bell inequality has been investigated in the
light of the $n$-particle Bell inequality discovered by Mermin
\cite{mermin}. However, for a bipartite system, it seems that some
important properties hidden behind the equivalence between
``entangled states'' and ``states that violate the Bell
inequality'' have not yet been discussed throughly. In this Letter,
based on Bell inequality and noncommutativity of measure
operators, we present an alternative definition of quantum entanglement
for two-qubit system; a uniform measure quantifying the degree of
entanglement is also proposed. One finds that, for pure states, to
obtain the maximal violations of Bell inequality, the measure
operators must satisfy some proper commutative relations. More
precisely, for a factorizable states, to obtain the maximum value
of inequality, the measures on one of the two particles must be
commutative; but for all entangled pure states (states are not
factorizable), the measures on each particle must be noncommutative.
This commutativity of measures is essential and important, and should
hold for any bipartite state.

Let us first briefly review the result of a pair of spin-$1/2$
particles. We perform the measurements either $A$ or $A^{\prime }$
on one particle, and either $B$ or $B^{\prime }$ on
the other, where $A,A^{\prime },B$ and $B^{\prime }$ denote any
physical variables with maximum absolute value $1$. Let $E(A,B),$
denote the expectation values of the product $AB$. After defining
$F$ as
\begin{equation}
F=E(A,B)+E(A,B^{\prime })+E(A^{\prime },B)-E(A^{\prime },B^{\prime
}), \label{ssdd}
\end{equation}
the Clauser-Horne-Shimony-Holt (CHSH) inequality can be
written as
\begin{equation}
-2\leq F\leq 2.  \label{ss}
\end{equation}
It holds in any theory of local hidden variables, but all
entangled pure states violate it. It has been shown by Gisin
\cite{gqn3} that the maximum value of the CHSH inequality for a
pure state $| \psi \rangle$ of a spin-$1/2$ particle pairs,
$F_{\max }(\psi )=2\sqrt{1+P_E^2},$ in which $P_E$ is a real
number between $0$ and $1$ -- zero for a factorizable pure states
and unit for a maximally entangled states$.$ For pure states, the
distinction between entangled states and disentangled states can
be equivalently regarded as whether the state can violate the
above inequality (\ref{ss}). Hence the number $P_E$ describes how
much a state violate this inequality.
Because the maximum value of the CHSH inequality of factorizable
states is $
2 $ that does not violate the inequality, these states are
disentangled, but for any other pure states $F_{\max }(\psi )>2,$
so they are entangled. Besides the equivalence between ``entangled
states'' and ``states that violate the Bell inequality'', there
exists another substantial difference between entangled states and
disentangled states: the commutativity of measures to realize the
maximum value of CHSH inequality are different for entangled states
and disentangled states.

It can be proved that for an entangled state, to obtain the
maximum of $F$ the measures on each particle must be
noncommutative, i.e., $[A,A^{\prime }]\neq 0$ and $[B,B^{\prime
}]\neq 0;$ however, for a disentangled states, to obtain the maximum of
$F$ the measures on one of the two particles must be commutative,
i.e., $[A,A^{\prime }]=0$ or $[B,B^{\prime }]=0.$ For notational
convenience, we assume a pure state
\begin{equation}
| \psi \rangle =k_1\left| 01\right\rangle +k_2\left|
10\right\rangle . \label{state}
\end{equation}
Let the observables  $A={\vec \sigma}\cdot{\vec n}$, $A^{\prime
}={\vec \sigma } \cdot {\vec n}^{\prime }$, $B={\vec
\sigma}\cdot{\vec m}$, $B^{\prime }={\vec \sigma } \cdot {\vec
m}^{\prime }$ be the spin projections onto
unit $3$-vectors $\vec n,\vec n^{\prime }$ for particle $a,$ and $\vec m,%
\vec m^{\prime }$ for particle $b$. A straightforward computation
yields
\begin{equation}
\left\langle \vec n\cdot \vec \sigma \otimes \vec m\cdot \vec \sigma
\right\rangle =-2k_1k_2(n_1m_1+n_2m_2)-n_3m_3,
\end{equation}
for all $3$-vectors $\vec n,\vec m.$ After some elaborations, we find that
when $n_2=n_2^{\prime }=m_2=m_2^{\prime }=0,$ $n_1=-\frac{k_1k_2}{|k_1k_2|}%
,\;n_3=0;\;n_1^{\prime }=0,\;n_3^{\prime }=-1;\;m_1=m_1^{\prime
}=2|k_1k_2|(1+4k_1^2k_2^2)^{-1/2},$ $m_3=-m_3^{\prime }=$ $%
(1+4k_1^2k_2^2)^{-1/2},$ the CHSH inequality is maximal for $|\psi
\rangle$, $ F_{\max }(\psi )=2\sqrt{1+4k_1^2k_2^2}$
\cite{gqn3,depure}$.$ Obviously, for
entangled states, i.e., $k_1\neq 0\neq k_2$, we have $[A,A^{\prime }]=-2i%
\frac{k_1k_2}{|k_1k_2|} {\vec \sigma} \cdot \hat e_2\neq 0,$
$\left[ B,B^{\prime }\right] =-8i \frac{|k_1k_2|}{1+4k_1^2k_2^2}
{\vec \sigma}\cdot \hat e_2\neq 0$, where $\hat e_2$ is the
direction vector. For instance, for the Bell state $|\psi ^{\pm
}\rangle=\frac{1}{ \sqrt{2}}\left( \left| 01\right\rangle \pm
\left| 10\right\rangle \right) ,$ one has $[A,A^{\prime }]=\mp 2i
{\vec \sigma}\cdot \hat e_2,$ $\left[ B,B^{\prime }\right] =-2i
{\vec \sigma}\cdot \hat e_2.$ However, for the factorizable states
, i.e., for $k_1=0$ or $k_2=0,$ we have $\left[ B,B^{\prime
}\right] =0,$ the measures on particle $b$ are commutative.

The relation between entanglement of states and the commutativity
of measures is very essential, which is an important evidence for
nonlocal quantum correlations. It has no reason to think that for
mixed states this evidence for nonlocal correlation can be
different from the one for pure states. Thus we come to the
uniform definition of entanglement for any state of bipartite
system based on the commutativity of measures for maximum value of
Bell inequality.

{\bf Definition}: {\it A state is said to be entangled, if the maximum
of CHSH expectation value $F_{\max }$ is obtain by noncommutative
measures on each particle of two particles. A state is said to be
disentangled, if the maximum of CHSH expectation value $F_{\max }$
is obtained by commutative measures on one of the pair particles.}

In order to quantify and measure the entanglement defined above, we
introduce another quantum expectation value $G$%
\begin{equation}
G=2E(C,D),  \label{gg}
\end{equation}
where $C=\vec \sigma \cdot \vec l$ , $D=\vec \sigma \cdot \vec h$ , and $%
\vec l,\vec h$ are two unit $3$-vectors. One can easily find that $G$ is a
CHSH expectation value by controlling the measures on one of the two
particles commutative$.$ In analog to the above discussion of the extreme
value of CHSH\ inequality, on finds for all the pure states the maximum
values of $G$ are the same, namely, $G_{\max }(\psi )=2$. For example, the
pure state as shown in Eq. (\ref{state}), $G(\psi
)=-4k_1k_2(n_1m_1+n_2m_2)-2n_3m_3,$ one easily proves that $G_{\max
}(\psi )=2$ when $n_1=m_1=n_2=m_2=0,$ $n_3=-m_3.$

Because $G$ is one of CHSH expectation values, so $F_{\max }(\psi )\geq
G_{\max }(\psi ).$ Only for disentangled states we have $F_{\max }(\psi
)=G_{\max }(\psi ),$ but for any entangled states $F_{\max }(\psi )>G_{\max
}(\psi ).$ This relation between the two quantum expectation values
$F$ and $G$ is
equivalence with the above definition of entanglement. We define the degree
of entanglement of a state $\rho $ as
\begin{equation}
P_E=\sqrt{\left( \frac{F_{\max }(\rho )}2\right) ^2-\left( \frac{G_{\max
}(\rho )}2\right) ^2.}  \label{degree}
\end{equation}
For the pure state, $2\leq F_{\max }(\psi )\leq 2\sqrt{2},$
$G_{\max }(\psi )=2$, so $0\leq P_E\leq 1.$ For example, $|\psi
\rangle =k_1\left| 01\right\rangle +k_2\left| 10\right\rangle $
and $|\phi \rangle =k_1\left| 00\right\rangle +k_2\left|
11\right\rangle ,$ one can easily obtain $P_E=2|k_1k_2|\leq 1.$
For the Bell states $|\psi ^{\pm }\rangle=\frac 1{\sqrt{2}}(\left|
01\right\rangle \pm \left| 10\right\rangle )$ and $|\phi ^{\pm
}\rangle=\frac 1{\sqrt{2}}(\left| 00\right\rangle +\left|
11\right\rangle ),$ $P_E=1.$ Obviously, when $k_1=0$ or $k_2=0,$
one has $P_E=0.$ This measure of entanglement is consistent
with the previous measures for pure states
\cite{myd9,depure,mdy12}. In the following, we extend our
discussion to any state of two qubits.

A state of two qubits can be completely described by the following
density matrix
\begin{equation}
\rho _{ab}=\frac 14({\bf I}_2\otimes {\bf I}_2+\vec \sigma ^a\cdot \vec u%
\otimes {\bf I}_2+{\bf I}_2\otimes \vec \sigma ^b\cdot \vec v{\bf +}%
\sum\limits_{i,j}^3\beta _{ij}\sigma _i^a\otimes \sigma _j^b),
\label{statea}
\end{equation}
where $\vec u$ and $\vec v$ are Bloch vectors for particles $a$ and $b$,
respectively; $\beta _{ij}$ are nine real numbers and can be regarded as the
components of a $9$-vector $\vec \beta ,$ which we call the correlation
vector; $\vec \sigma ^{a,b}$ are Pauli matrices and ${\bf I}_2$ is $2\times
2 $ unit matrix. By making partial trace, the reduced density matrices of
two particles can be obtained as,
$\rho _a={\rm Tr}_b(\rho _{ab})=({\bf I}_2
+{\vec \sigma} ^a\cdot {\vec u})/2$ and
$\rho _b={\rm Tr}_a(\rho _{ab})=
({\bf I}_2 +{\vec \sigma}^b \cdot {\vec v})/2$.

We then have CHSH expectation $F$ of the state $\rho_{ab}$ as
\begin{equation}
F(\rho _{ab})={\rm Tr}[\rho _{ab}(A\otimes B+A\otimes B^{\prime }+A^{\prime }\otimes
B-A^{\prime }\otimes B^{\prime })].  \label{fr}
\end{equation}
After introducing
\begin{equation}
T_{ij}=(n_i+n_i^{\prime })m_j+(n_i-n_i^{\prime })m_j^{\prime },\;\;i,j=1,2,3,
\label{tt}
\end{equation}
which are components of a $9$-vector $\vec T$, one obtains
\begin{equation}
F(\rho _{ab})=\sum\limits_{i,j}\beta _{ij}T_{ij}={\vec \beta} \cdot{\vec T},
\end{equation}
which is just the inner product of two $9$-vectors $\vec \beta $
and $\vec T$. For any $\vec T{\bf (}n,n^{\prime },m,m^{\prime }),$
$|\vec T|=
\sqrt{ \sum_{i,j}T_{ij}^2 }=2$, so that
$F(\rho _{ab})=2| {\vec \beta} |\cos [\theta (\vec \beta ,\vec T{\bf )],}$
in which $|{\vec \beta}
|=\sqrt{ \sum_{i,j}\beta _{ij}^2 }$ is the length of the vector
$\vec \beta $, $\theta (\vec \beta ,\vec T{\bf )}$
is the angle between ${\vec \beta}$ and ${\vec T}$. We denote the
minimum angle of $\theta (\vec \beta ,\vec T{\bf )}$ as
$\gamma =\min \{\theta (\vec \beta ,\vec T{\bf %
)\},}$ we then have $F_{\max }(\rho _{ab})=2|{\vec\beta} |\cos \gamma .$

Similarly, we get $G(\rho _{ab})=2|\vec \beta |\cos [\theta (\vec
\beta ,\vec D{\bf )],}$ and $G_{\max }(\rho _{ab})=2|\beta |\cos \delta ,$
where $\vec D$ is a vector with components $D_{ij}=2l_ih_j$ and $\delta
=\min \{\theta (\vec \beta ,\vec D{\bf )}\}.$ Consequently, we can formally
express the degree of entanglement for a state $\rho _{ab}$ as
\begin{equation}
P_E=|\vec \beta |\sqrt{\cos ^2(\gamma)-\cos ^2(\delta)}.
\label{pe1}
\end{equation}

For a given state $\rho _{ab}$, one can always find the vectors $\vec T$
and $\vec D$ such that $F$ and $G$ reach their maximum values, respectively.
This implies
that we obtain proper relations among $\vec n,\vec n^{\prime },\vec m,$ and
$\vec m^{\prime }$, so do $\vec l$ and $\vec h.$ At this moment, we
have $[A,A^{\prime
}]=2i{\vec \sigma} \cdot(\vec n\times \vec n^{\prime })=2i{\vec
\sigma} \cdot\vec X$ and
$[B,B^{\prime }]=2i{\vec \sigma}\cdot(\vec m\times \vec m%
^{\prime })=2i{\vec \sigma} \cdot\vec Y.$ Denoting the angle between $%
\vec T$ and $\vec D$ (which make $F$ and $G$ reach their maximum
values, respectively) is $\eta ,$ one then obtains
\begin{equation}
|\vec X|\cdot |\vec Y|=\frac{4G_{\max }\cdot P_E}{(F_{\max
})^2}=\sin (2\eta ). \label{geo}
\end{equation}
This formula shows an important relation between commutativity of measures
and the extreme value of $F$. Obviously, this relation is only determined by
the direction of $\vec \beta .$ For $F_{\max }=G_{\max }$,
i.e., the measures on one of two particles are commutative, one must have
$|{\vec X}|=0$ or $|{\vec Y}|=0$, from the above formula we obtain $P_E=0$.
This feature is in agreement with Eq. (\ref{degree}) as well as our
definition of entanglement.

What does it mean when the entanglement degree of a state is zero?
We can find that there are only two cases when $P_E=0.$ Case 1: when
$|\vec \beta
|=0,$ $F_{\max }(\rho )=G_{\max }(\rho )=0,$ so does $P_E.$ Case 2: when $%
\beta _{ij}=w_ir_j,$ where $\vec w$ and $\vec r$ are two $3$-vectors.
It is easy to prove that $F_{\max }(\rho )=2|\vec w|\cdot |\vec r
| $ and $G_{\max }(\rho )=2|\vec w|\cdot |\vec r|,$ so that
$P_E=0$. For these two cases, one finds that the rank of the
so-called correlation matrix
\[
\beta _M=\left(
\begin{array}{ccc}
\beta _{11} & \beta _{12} & \beta _{13} \\
\beta _{21} & \beta _{22} & \beta _{23} \\
\beta _{31} & \beta _{32} & \beta _{33}
\end{array}
\right)
\]
is not higher than one, namely, ${\rm rank}[\beta _M]\leq 1.$ It has been
proved that, for ${\rm rank}[\beta _M]\leq 1$, a state $\rho $ is
separable, i.e., this state can be represented as a mixture of factorizable
pure states \cite{hans}. In other words, for ${\rm rank}[\beta _M]\leq 1$
and $|\vec \beta |\neq 0,$ one can always find two unit $3$-vectors
$\vec w$ and $
\vec r$ by which the components of correlation vector are expressed as $%
\beta _{ij}=w_ir_j,$ this must lead to $P_E=0.$ Then we come to an
important result: for a state of bipartite system $\rho _{ab}$, if
and only if ${\rm rank}[\beta _M]\leq 1,$ its degree of
entanglement is zero, i.e., $P_E=0;$ at this moment the state is
separable$.$ In other words, if the rank of correlation matrix of
a state is higher than $1$, then this state is entangled with
nonzero degree of entanglement.

The above analysis implies that not all of the separable state has zero
degree of entanglement. To see this point clearly, let us consider
the Werner state \cite{werner} in the form
\begin{equation}
\rho =(1-\alpha )\frac{I_4}4+\alpha \left| \psi ^{+}\right\rangle
\left\langle \psi ^{+}\right| .  \label{wer}
\end{equation}
Its correlation matrix of the above state is $\beta _M=\left(
\begin{array}{ccc}
\alpha & 0 & 0 \\
0 & \alpha & 0 \\
0 & 0 & -\alpha
\end{array}
\right) .$ It has been proved that when $\alpha \leq \frac 13$ the
above state is separable \cite{sep}. For the
maximally entangled state $|\psi ^{+} \rangle=\frac
1{\sqrt{2}}(\left| 01\right\rangle +\left| 10\right\rangle )$, its
corresponding correlation matrix is $\beta _M=\left(
\begin{array}{ccc}
1 & 0 & 0 \\
0 & 1 & 0 \\
0 & 0 & -1
\end{array}
\right) $; obviously this correlation vector is paralleling to the
correlation vector of the state given in (\ref{wer}), so the
commutative relations of measures on each particle for these two
state should be the same, thus the state given in (\ref{wer}) is
entangled based on our definition. From (\ref{pe1}) it can be
easily obtain $P_E=\alpha$. For the above state, although it is
saparable for $\alpha \leq 1/3$, it still has nonzero degree
of entanglement $P_E=\alpha $.

Therefore, the degree of entanglement of a state is zero means that the
state is separable. However, a separable state does not always
mean that its degree of entanglement is zero (based on our
definition) unless ${\rm rank}[\beta _M]\leq 1$. This viewpoint is
somewhat different from other considerations \cite{myd9,h3,jens} in
which a separable state is regarded as a disentangled state.

The above discussions of entanglement can be understand in another way. We
assume an inequality for any state as
\begin{equation}
-|G_{\max }(\rho )|\leq F(\rho )\leq |G_{\max }(\rho )|.  \label{fu}
\end{equation}
Then, one can find that any entangled state must violate this
inequality. The maximum violations of the above inequality can be
formally expressed as
\begin{equation}
F_{\max }(\rho )=2\sqrt{\left( \frac{G_{\max }(\rho )}2\right) ^2+P_E^2}.
\label{dddd}
\end{equation}
For pure states, the inequality (\ref{fu}) reduces to Eq. (\ref{ss}).
Obviously, Eq. (\ref{dddd}) is just a straightforward
generalization of the formulae that Gisin presented for pure
states \cite{gqn3}.

In conclusion, we have presented an alternative definition of quantum
entanglement for any state of bipartite system based on the CHSH
inequality. In our definition, if a state is disentangled, the
maximum of the CHSH expectation value can be only obtained by
commutative measures on one of the two particles. Meanwhile,
this state is separable with the rank of correlation matrix ${\rm rank}
[\beta _M]\leq 1$. Under this definition, we introduce a uniform measure
to quantify the entanglement (see Eq.(\ref{dddd})), which is a
straightforward extension for the case of pure state. We also find
that there exists an important relation between the commutativity of
measures and maximum of CHSH expectation value, which exhibits a
simple geometric significance (see Eq. (\ref{geo})).

It can be noted that our definition of entanglement is different
from the traditional understanding of entanglement for mixed state
of bipartite system. In our definition not all the separable
states are disentangled, only the states with ${\rm rank}[\beta
_M]\leq 1$ are disentangled; but in traditional consideration, the
separable states should be disentangled. We cannot tell which
definition is good since the entanglement of mixed state has
not yet been resolved well. Anyway, whether there exist entanglement
and nonlocal correlations in separable states is still a debate so
far \cite{oc1,oc2}. Nevertheless, the understanding of quantum
entanglement based on our definition has a clear and simple
physical picture. The distinction between entangled states and
disentangled states is clear and in the uniform physical
interpretation for both pure and mixed states. The measure of
entanglement we have proposed makes the evaluation of entanglement for
bipartite system very easy. These features should facilitate the
discussion of some problems concerning entanglement and have
possible important applications in the future.

\end{document}